\documentclass[nofootinbib]{revtex4}
\usepackage[english]{babel}
\usepackage{array,booktabs}
\usepackage{array} 
\usepackage{lipsum}   
\usepackage{calc}
\usepackage{pdflscape}
\usepackage{color}
\usepackage{float}
\usepackage[font=small,labelfont=bf]{caption}
\usepackage{graphicx}
\usepackage{amsmath}
\usepackage[nodisplayskipstretch]{setspace}

\usepackage{setspace}
\usepackage{tabularx}

\usepackage{float}
\usepackage{color}
\usepackage{amsmath}
\usepackage{float}
\usepackage{calc}
\usepackage{pdflscape}
\usepackage{color}
\usepackage{float}
\usepackage{subfigure}
\usepackage[font=small,labelfont=bf]{caption}
\usepackage{graphicx}
\begin{document}{\setlength\abovedisplayskip{4pt}}

\title{Neutrino-Nucleon Cross section in Ultra High Energy Regime\footnote{Talk presented at XXI DAE-BRNS HIGH ENERGY PHYSICS SYMPOSIUM 08-12 DECEMBER 2014, IIT GUWAHATI, Assam, India}}
\author{Kalpana Bora\footnote{E-mail: kalpana.bora@gmail.com} , Neelakshi Sarma\footnote{E-mail: nsarma25@gmail.com}} 
\affiliation{Department Of Physics, Gauhati University, Guwahati-781014, India}


\begin{abstract}
Neutrino Physics is now entering precision era and neutrino-nucleon cross sections are an important ingredient in all neutrino oscillation experiments. Specially, precise knowledge of neutrino-nucleon cross sections in Ultra High Energy (UHE) regime (TeV-PeV) is becoming more important now, as several experiments worldwide are going to observe processes involving such UHE neutrinos. In this work, we present new results on neutrino-nucleon cross-sections in this UHE regime, using QCD.
\end{abstract}
\maketitle

\section{Introduction}	
\setlength{\baselineskip}{13pt}
Neutrino is one of the most interesting elementary particle that plays an important role in understanding nuclear as well as particle physics. Many properties of this particle have already been measured experimentally, still some of them are yet to be measured. Unknown quantities include mass heirarchy, CP violation phase, nature of neutrinos (whether Majorana or not), exact mass of neutrinos, whether neutrinos contribute to dark matter etc. Many experiments are ongoing/planned all over the world to measure these quantities. In any neutrino oscillation experiment, the neutrino is allowed to scatter off a nucleon/nucleus, and hence precise knowledge of neutrino-nucleon interaction cross section is required to do any analysis. This interaction cross section plays a vital role in different neutrino oscillation experiments at different stages of calculation. Neutrino interactions across various energy scales can be described below \cite {1}:\\
(a) Thresholdless process ($E_\nu$ $ \sim 0 - 1$ MeV)
\\When neutrino has zero momentum, such processes occur which include coherent scattering and neutrino capture on radioactive nuclei (enhanced or stimulated beta decay emission).
\\(b) Low energy nuclear process ($E_\nu$ $ \sim 0 - 100$ MeV)
\\At such energy scale it is possible to probe the target nucleus at smallest length scales.
\\(c) Intermediate energy process ($E_\nu \sim 0.1 - 20$ GeV)
\\Neutrino-nucleon scattering processes becomes complicated at such energy scale which can be classified mainly as elastic and quasi-elastic (QE) scattering. In such scattering processes neutrino scatters off an entire nucleon giving rise to a nucleon or multiple nucleons from the target nucleus. Charged current neutrino scattering process is referred to as `quasi-elastic scattering' whereas neutral current scattering is known as `elastic scattering'.
\\(d) High energy cross sections ($E_\nu \sim 20 - 500$ GeV)
\\Neutrino can resolve the individual quark constituents of the nucleon at such energy scale and this type of scattering is called Deep inelastic scattering (DIS) process.
\\(e) Ultra high energy (UHE) neutrino scattering ($E_\nu \sim 0.5$ TeV $- 1$ EeV)
\\The latest highest energy neutrino recorded so far is $\sim 10^{15}$ eV which provide an opportunity for researchers to work with UHE neutrinos from astrophysical sources \cite{1,2}. Some of the ongoing/planned experiments for observation of UHE neutrinos from astrophysical sources are Baikal \cite {3}, ANITA \cite {4}, RICE \cite {5}, AMANDA \cite {6}, HiRes \cite {7}, ANTARES \cite {8}, IceCube \cite {9} etc.

\section{Neutrino scattering at UHE regime}
In DIS, the neutrino scatters off a quark in the nucleon via the exchange of a virtual W or Z boson producing a lepton and a hadronic system in the final state \cite {10}. Both charged current (CC) and neutral current (NC) processes are possible like 
$\nu_{l} + N \rightarrow l^{-}$ + X and  $\bar{\nu}_l + N \rightarrow l^{+} + X .$   
\\The Feynman diagram for the process $\nu_{l} (p_\nu)+ N (p_{N})\rightarrow l^{-}$ $(p_{l})+ X(p_{X})$ is shown in Figure 1(a) and in the quark-parton model with elementary $W^{+}(q) + d(p_{i})\rightarrow u(p_{f})$ transition is shown in Figure 1(b). 
\begin{center}
\begin{figure}
\includegraphics[width = 8.7 cm]{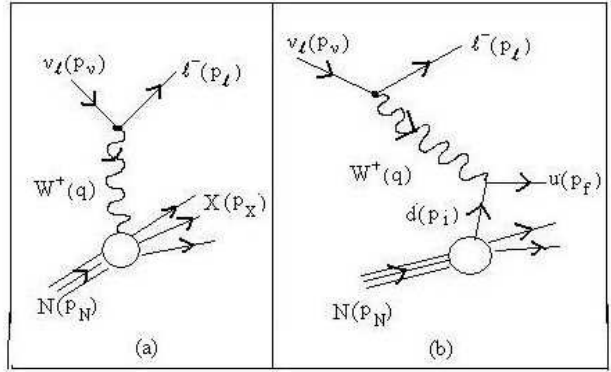}
\caption{(a) Diagram of the $\nu_{l} (p_\nu)+ N (p_{N})\rightarrow l^{-}$ $(p_{l})+ X(p_{X})$ charged-current DIS process. (b) Diagram of the same process in the quark-parton model.}
\end{figure}
\end{center}

\subsection{DGLAP evolution equation} 
The QCD improved proton structure function \cite {11} can be written as 
\begin{equation}
\frac{F_2(x, Q^2)}{x} = \sum_{q} e_{q}^{2} \int_{x}^{1}\frac{dy}{y}q(y)[\delta(1-\frac{x}{y})+ \frac{\alpha_s}{2\pi}P_{qq}(\frac{x}{y})log\frac{Q^2}{\mu^2}]
\end{equation}
\\where $\mathit{y}$ is the fraction of proton's momentum carried by any of its partons,  $q(y) = f_{q}(y)$ is the quark structure function inside the proton, $\mathit{x}$ is the Bjorken variable, $Q^{2}$ is the transverse momentum square of exchanged gauge boson and $P_{qq}(\frac{x}{y})$ is the splitting function which represents the probability of a quark emitting a gluon and becoming a quark with momentum reduced by a fraction $\mathit{z}$ (where $\mathit{z}$ = $\frac{x}{y}$) . In QCD, the $F_{2}$ structure function is a function of both $\mathit{x}$ and $Q^{2}$ , the variation with $Q^{2}$ being only logarithmic. The evolution of quark density function $q(x, Q^{2})$ with $\mathit{x}$ and $Q^{2}$ can be expressed as Altarelli-Parisi (AP) equation.
\begin{equation}
\frac{dq(x,Q^2)}{dlog{Q^2}} = \frac{\alpha_s}{2\pi}\int_{x}^{1}\frac{dy}{y}q(y, Q^{2})P_{qq}(\frac{x}{y})
\end{equation}
\subsection{Total cross section at UHE}
In neutrino-nucleon scattering, for the range of neutrino energies $10^{6} GeV \le E_{\nu} \le 10^{13} GeV,$ the value of $\mathit{x}$ lies in the range $10^{-2} \le x \le 10^{-9}$.  
The total charged current cross-sections for neutrino-nucleon scattering \cite {12} can be written as
\begin{equation}  
\sigma^{\nu N}_{CC}(E_{\nu}) = \int_{Q_{min}^{2}}^{s} dQ^{2 }\int_{\frac{Q^2}{s}}^{1}dx \frac{d^{2}\sigma_{CC}}{dx dQ^{2}}(E_{\nu}, Q^{2}, x) = \frac{G_{F}^{2}}{4\pi}\int_{Q_{min}^{2}=1}^{2mE_{\nu}} dQ^{2}(\frac{M_{W}^{2}}{Q^{2}+M_{W}^{2}})^{2}\int_{\frac{Q^{2}}{2mE_{\nu}}}^{1}\frac{dx}{x}(F_{2}^{\nu})
\end{equation} 
where $F_{2}^{\nu}$ is the neutrino-nucleon structure function, $\mathit{s} = 2mE_{\nu}$ where $\mathit{s}$ is the Mandelstam variable which is the total energy in the centre of mass frame , $\mathit{m}$ is the nucleon mass , $G_{F}$ is the Fermi constant and $M_{W}^{2}$ is the squared mass of intermediate W-boson . Similar expression can be obtained for neutral current cross section. For the flavor-symmetric ($q\bar{q}$)N interaction at $\mathit{x}\ll 0.1$, the neutrino-nucleon structure function, $F_{2}^{\nu}(x, Q^{2})$ can be related to electromagnetic structure function, $F_{2}^{p}(x, Q^{2})$ \cite {13} as 
\begin{equation}
F_{2}^{\nu}(x, Q^{2}) = \frac{n_{f}}{\sum_{q}^{n_{f}}Q_{q}^{2}} F_{2}^{p}(x, Q^{2})
\end{equation}
where $n_{f}$ is the number of flavors and $Q_{q}$ is the quark charge. The neutrino-nucleon scattering process involves scattering of neutrino off quarks as shown in Figure 1(b). Thus, the knowledge of proton structure function can be utilised in calculation of neutrino-proton scattering cross-section. This idea has been utilised in the present work. In the following section, we discuss the behaviour of $F_{2}^{p}$ in DAL (Double Asymptotic Limit) which in turn would be used to calculate neutrino-nucleon scattering cross-section in UHE regime.
\section{Double Asymptotic Limit (DAL) and UHE Neutrino Cross section}
In ($\textit{e-p}$) DIS, in the next to leading order, scaling violations occur through gluon bremsstrahlung from quarks and quark pair creation from gluons. At small $ x< 10^{-2}$, the latter process dominates the scaling violations. This property can be exploited to extract gluon density from the slope $\frac{dF_{2}}{dlnQ^{2}}$ of the proton structure function. The general equations \cite{14} describing the $Q^{2}$ evolution of the quark density and gluon density respectively are
\begin{equation}
\frac{dq^{i}(x,t)}{dt} = \frac{\alpha(t)}{2\pi}\int_{x}^{1}\frac{dy}{y}[\sum_{j=1}^{2f}q^{j}(y,t)P_{qq}(\frac{x}{y}) + G(y,t)P_{qG}(\frac{x}{y})] 
\end{equation}
\begin{equation}
\frac{dG(x,t)}{dt} = \frac{\alpha(t)}{2\pi}\int_{x}^{1}\frac{dy}{y}[\sum_{j=1}^{2f}q^{j}(y,t)P_{Gq}(\frac{x}{y}) + G(y,t)P_{GG}(\frac{x}{y})]
\end{equation}
where $P_{qq}(\frac{x}{y}), P_{qG}(\frac{x}{y}), P_{Gq}(\frac{x}{y}), P_{GG}(\frac{x}{y})$ are the splitting functions and $t = ln\frac{Q^{2}}{Q_{0}^{2}}$. Assuming that the quark densities are negligible and the non-singlet contribution $F_{2}^{NS}$ can be ignored safely at small $\mathit{x}$ in AP equation, for $F_{2}$, the equation becomes
\begin{equation}
\frac{dF_{2}(x, Q^{2})}{dlnQ^{2}} = \frac{10\alpha_{s}}{9\pi}\int_{x}^{1}dx^{'}P_{qg}(x^{'})\frac{x}{x^{'}}g(\frac{x}{x^{'}},Q^{2})
\end{equation}
Here $xg(x,Q^{2}) = G(x, Q^{2})$ is the gluon momentum density and $g(x,Q^{2})$ is the gluon number density of the proton and $\frac{x}{x^{'}} = \frac{x}{y}$. Rearranging equation (7) we have 
\begin{equation}
\frac{dF_{2}(x, Q^{2})}{dlnQ^{2}} = \frac{5\alpha_{s}} {9\pi}\int_{x}^{1}dy\frac{x}{y}g(y, Q^{2})\frac{1}{y^{2}}[x^{2}+(y-x)^{2}]
\end{equation}
Substituting $y = \frac{x}{1-z}$ we can write RHS of equation (8) as 
\begin{equation}
\frac{5\alpha_{s}}{9\pi}\int_{0}^{1-x} dz G(\frac{x}{1-z}, Q^{2})[z^{2}+(1-z)^{2}]
\end{equation}
Expanding $G(\frac{x}{1-z},Q^{2})$ about $z = \frac{1-x}{2}$ and keeping terms upto the first derivative of G in the expansion we have
 
\begin{equation}
 G(\frac{x}{1-z},Q^{2})= G(\frac{2x}{1+x},Q^{2}) + (z-\frac{1-x}{2})\frac{4x}{(1+x)^{2}}\frac{dG(x^{''},Q^{2})}{dx^{''}} \pmb{\Bigg\vert}_{x^{''}=\frac{2x}{1+x}}
\end{equation}
 
When this expansion is used in equation (8) we get 
\begin{equation}
\frac{dF_{2}(x,Q^{2})}{dlnQ^{2}} = \frac{5\alpha}{9\pi} \frac{(A+Ax+2B)^{2}}{(1+x)(A+Ax+4B)}G(y^{'},Q^{2})
\end{equation}
where $y^{'} = [\frac{2x}{1+x}\frac{(A+Ax+4B)}{(A+Ax+2B)}]$, $A = [\frac{2(1-x)^{3}}{3}-(1-x)^{2}+(1-x)]$, and $B = [\frac{(1-x)^{4}-(1-x)^{3}}{6}]$ . \\ In the limit $x\rightarrow 0$ equation (11) reduces to 
\begin{equation}
\frac{dF_{2}(x,Q^{2})}{dlnQ^{2}} = \frac{10\alpha_{s}}{9\pi} \frac{(1-x)^{2}}{(1-1.5x)} G (2x\frac{(1-1.5x)}{(1-x^{2})}, Q^{2})
\end{equation}
Using the above, double asymptotic expression \cite{14} for $F_{2}$ in small $\mathit{x}$ and large $Q^{2}$ (DAL) limit, we can write
\begin{equation}
F_{2}^{p} \sim \frac{exp\sqrt{\frac{144}{33-2n_{f}}\xi ln(\frac{1}{x_{1}})}}{(\frac{144}{33-2n_{f}}\xi ln(\frac{1}{x_{1}}))^{\frac{1}{4}}}
\end{equation}
with $\xi = ln(\frac{ln\frac{Q^{2}}{\Lambda ^{2}}}{ln\frac{Q_{0}^{2}}{\Lambda ^{2}}})$, 
$x_{1} = \frac{2x-3x^{2}}{1-x^{2}}$ , $n_{f}$ is the number of flavors, $Q_{0}^{2}$ is the value at which the input parton parameterization is to be used and $\Lambda$ is the QCD mass scale. $F_{2}^{p}$ in equation (13) in DAL can be parametrized as 
\begin{equation}
F_{2}^{p} \sim x^{-\lambda(Q^{2})}
\end{equation}
This behaviour of $F_{2}^{p}$, when used in equation (3) gives
\begin{equation}
\sigma^{\nu N}_{CC}(E_{\nu})\approx \frac{G_{F}^{2}}{4\pi}\int_{Q_{min}^{2}=1}^{2mE_{\nu}} dQ^{2}(\frac{M_{W}^{2}}{Q^{2}+M_{W}^{2}})^{2}\int_{\frac{Q^{2}}{2mE_{\nu}}}^{1}\frac{dx}{x}(x^{-\lambda(Q^{2})})
\end{equation}
where $\lambda(Q^{2}) = a - b.e^{-cQ^{2}}$ and the values of constants are found to be as $a = 0.486 , b = 0.272$ and $c = 0.002.$ From equations (14) and (15) it can be seen that 
$$\sigma_{CC}^{\nu N} = A \frac{G_{F}^{2}}{4\pi}\int_{Q_{min}^{2}=1}^{2mE_{\nu}} dQ^{2}(\frac{M_{W}^{2}}{Q^{2}+ M_{W}^{2}})^{2}  (\frac{x^{-\lambda(Q^{2})}}{-\lambda(Q^{2})})$$ 
\begin{equation}
= A \frac{G_{F}^{2}}{4\pi}\int_{Q_{min}^{2}=1}^{2mE_{\nu}} dQ^{2}(\frac{M_{W}^{2}}{Q^{2}+ M_{W}^{2}})^{2} 
(\frac{x^{-a + b.e^{-c.Q^{2}}}}{b.e^{-c.Q^{2}}-a})
\end{equation}
in low $\mathit{x}$ and high $Q^{2}$ regime. Here A is normalisation constant. \\ 
Further analysis is in progress \cite{15}.

\section{Summary}
To summarize, in this work we have utilized the DAL behaviour of proton structure function in calculating the neutrino scattering cross section (see equation (16)). As DAL is applicable in low - $\mathit{x}$ and high $Q^{2}$ regime we expect that this behaviour could be applicable for explaining UHE neutrino scattering cross section in this regime. 


\end{document}